\documentclass[3p, twocolumn]{elsarticle}

\usepackage{graphicx}  
\usepackage[english]{babel}
\usepackage{blindtext}
\usepackage[noams]{SIunits} 
\usepackage{booktabs}

\usepackage{amsmath}   

\begin{document}

\title{OTA based 200~G$\Omega$ resistance on 700~$\mu m^2$ in 180~nm CMOS for neuromorphic applications}
\author[ini]{C.Mayr\corref{cor1}}
\ead{cmayr@ini.uzh.ch}
\author[pmd]{M.Schultz}
\ead{m.schultz@pmdtec.com}
\author[tud]{M. Noack}
\ead{noack@iee.et.tu-dresden.de}
\author[tud]{S.Henker}
\ead{henker@iee.et.tu-dresden.de}
\author[tud]{J. Partzsch}
\ead{partzsch@iee.et.tu-dresden.de}
\author[tud]{R. Sch\"uffny}
\ead{schuffny@iee.et.tu-dresden.de}

\cortext[cor1]{corresponding author}

\address[ini]{Institute of Neuroinformatics, University of Zurich and ETH Zurich, Switzerland}
\address[tud]{Institute of Circuits and Systems, University of Technology Dresden, Helmholtzstrasse 18, 01069 Dresden, Germany}
\address[pmd]{pmdTechnologies GmbH, Am Eichenhang 50, 57076 Siegen, Germany}

\begin{abstract}
Generating an exponential decay function with a time constant on the order of hundreds of milliseconds is a mainstay for neuromorphic circuits. Usually, either subthreshold circuits or RC-decays based on transconductance amplifiers are used. In the latter case, transconductances in the 10~pS range are needed. However, state-of-the-art low-transconductance amplifiers still require too much circuit area to be applicable in neuromorphic circuits where $>$100 of these time constant circuits may be required on a single chip. We present a silicon verified operational transconductance amplifier that achieves a g$_m$ of 5~pS in only 700~$\mu m^2$, a factor of 10-100 less area than current examples. This allows a high-density integration of time constant circuits in target appliations such as synaptic learning or as driving circuit for neuromorphic memristor arrays.
\end{abstract}

\maketitle

\section{Introduction}
\label{intro}

Neuromorphic circuits that operate in real time have to exhibit the same timescale as biological neurons \cite{chou10}. This means a circuit has to generate multiple exponential decay functions with time constants of up to hundreds of milliseconds, employed e.g. as the membrane time constant, presynaptic adaptation or postsynaptic current trace \cite{rolls10}. 

In the form of the gm-C configuration \cite{koickal07}, Operational Transconductance Amplifiers (OTA) are widely used to produce exponential decay functions. To achieve time constants in the above range, large capacitances and low transconductances are needed. Circuit area limits capacitances to $\leq$1~pF, requiring transconductances on the order of 10~pS \cite{hu11,veeravalli02}.

While some papers have reported such low values, their large circuit area implementations prohibit an integration of multiple instances of these OTAs for each neuron circuit \cite{villegas11}. Another common problem is that the low transconductance is achieved through a switching scheme \cite{chou10,villegas11}, requiring complex state machines \cite{noack12} and dedicated clock generators \cite{eisenreich09} and foregoing the asynchronous paradigm of neuromorphic circuits \cite{Indiveri2006}.

In this paper we present a transconductance amplifier designed in 180~nm CMOS that is targeted at producing exponentially decaying voltage traces for use in a novel learning rule \cite{mayr10a} and for driving a particular type of memristor \cite{mayr12b}. In keeping with the neuromorphic tenet, the OTA is time-continuous, allowing asynchronous triggering for the waveform generation.

Especially for future nanoscale memristor arrays, driver circuits are needed that can be integrated at very high density in the CMOS substrate underlying the memristor array. By employing a combination of circuits techniques from literature, we arrive at an OTA that achieves a transconductance on par with the lowest reported, while occupying a factor of 10-100 less circuit area than previous work. This agressive scaling incurs some penalty in terms of linearity compared to literature. However, we show that this linearity reduction is not critical  for the two specific applications above, which rely only on reasonable replication of an exponential curve shape replication.

\section{Circuit}
\label{sec_circuits}

The design target for the low transconductance OTA can be summarized as follows: the OTA has to exhibit a wide voltage swing at input and output, thus allowing a large voltage range for the neuron state variables and memristor drive waveforms. As it is primarily built for replicating those state variables and not for e.g. a linear filter \cite{villegas11}, OTA linearity can be somewhat neglected in favor of integration density. In order to minimize deviations caused by process variations, the use of transistors working in the weak inversion region has to be avoided where possible. 

\begin{figure}[!htbp]
\begin{center}
\includegraphics[width=3.0in]{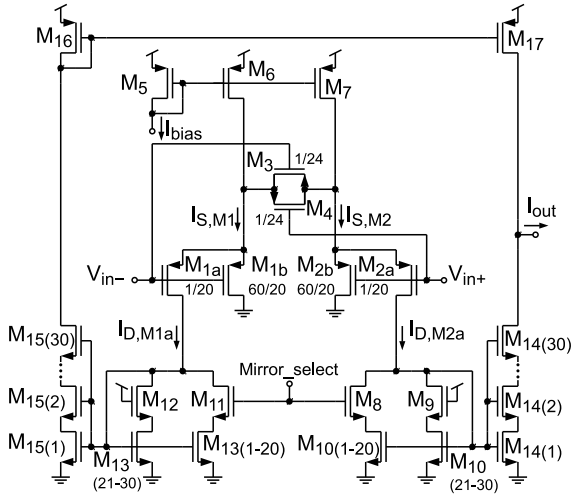}
\caption{Transconductance amplifier with series-parallel current mirrors, current splitting and source degeneration. Numbers are $W/L$ for each transistor. }
\label{ota_circuit}
\end{center}
\end{figure}

A combination of three techniques is used to achieve very low transconductance in a compact area: current splitting, source degeneration and current downscaling via extremely sized current mirrors. Figure \ref{ota_circuit} shows the resulting OTA.

The input differential pair M1 and M2 in Fig. \ref{ota_circuit} has an overall $\frac{W}{L}$ ratio of $\frac{61}{20}$. It uses current splitting \cite{veeravalli02} 
 with a splitting ratio of $N_{split}=\frac{I_{D,M1a}}{I_{S,M1}}=\frac{I_{D,M2a}}{I_{S,M2}}=\frac{1}{61}$. This allows to operate the tail current source and the differential pair in strong inversion while having an effective transconductance significantly smaller than that of a conventional differential pair. Operation in strong inversion also extends the input voltage swing compared to subthreshold circuits \cite{razavi2001}. Transistors M3 and M4 perform negative feedback to improve linearity and esp. to further lower the transconductance \cite{sanchez00}. Due to their low drain-source voltage, M3 and M4 operate in the linear region. 
$I_{D\text{,M1a}}$ and  $I_{D\text{,M2a}}$ are each fed into a  modified series-parallel current mirror \cite{arnaud03} with a switchable current scaling factor  between $N_{CScale}=\frac{I_{D,M15}}{I_{D,M1a}}=\frac{I_{D,M14}}{I_{D,M2a}}=\frac{1}{900}$ and  $N_{CScale}=\frac{1}{300}$. Due to the low current values after current splitting, the series-parallel current mirror (M10, M13--M15) as well as the output current mirror (M16, M17) operate in the weak inversion region. 
Equation \ref{eq:gmtotal} describes the overall small signal transconductance.  
 \begin{equation}
g_{m, total}= N_{split}N_{CScale} \frac {g_{m1}}{1+\frac{\beta_{1}}{4 \beta_{3}} }
\label{eq:gmtotal}
\end{equation}   
 
The transconductance of the source degenerated differential pair is computed according to \cite{sanchez00}, with the transconductance $g_{m1}$ of M1 downscaled by the $\frac{W}{L}$ ratio of M1 and M3 ($\beta_{1}$ resp. $\beta_{3}$). This transconductance is multiplied by current scaling and splitting factors. To keep M1 - M7 in strong inversion the circuit was designed to operate from $I_{bias}= 1$~$\mu A$ to $I_{bias}= 10$~$\mu A$. The combination of this $I_{bias}$-range plus the selectable current mirror allows a range of (nominal) 5 to 50~pS for the overall transconductance $g_{m, total}$. The power dissipation is governed by $I_{bias}$ and the supply voltage $V_{DD}$, it equates to:
\begin{equation}
P_{diss} = 3 I_{bias} \cdot V_{DD}
\end{equation}

The output current of the current mirrors M14 and M15 is neglected because of the large current mirror scaling factor.

A layout was carried out with $0.5\times 0.5$~$\mu m^2$ unity transistors, resulting in an overall chip area of  $700$~$\mu m^2$. The layout was designed in a strict common centroid regime aided by the large number of unity transistors. Additional dummy transistors were placed in the surrounding area and the propositions of \cite{arnaud06} were applied. 

\begin{figure}[!htbp]
\begin{center}
\includegraphics[width=3.0in]{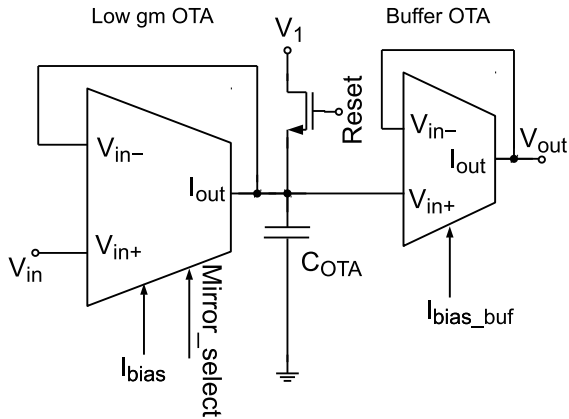}
\caption{Transconductance amplifier in OTA-C configuration with decoupling OTA buffer that separates pad from OTA-C capacitance. }
\label{fig_ota_c}
\end{center}
\end{figure}

On the chip, the OTA is integrated in a gm-C configuration \cite{arnaud06}, see Fig. \ref{fig_ota_c}. That is, the OTA operates in a feedback loop with the output connected to a gate-capacitance, allowing to replicate exponential RC-curves via a square waveform input at V$_\mathrm{in}$. The V$_1$ input can be used to characterize the measurement chain without the OTA. In addition, this allows to reset C$_\mathrm{OTA}$, enabling waveforms with an instantaneous onset and exponential decay. The buffered voltage signal can e.g. be used directly to drive voltage-dependent memristors, with voltage levels and waveforms already compatible with the memristors of \cite{mayr12b}. For use as a postsynaptic current (PSC) waveform in the plasticity rule of \cite{mayr10b}, the OTA-buffer at the left side of Fig. \ref{fig_ota_c} can be operated without feedback, acting as a voltage to current converter configurable by its bias current.

\section{Noise}
\label{sec_noise}
To judge the performance of the OTA with respect to neuromorphic exponential signal generation, an estimate of the achievable signal-to-noise ratio is necessary.
Thus, this section discusses possible noise sources and gives simulated and measured results for the OTA noise.
In the following, the noise performance of the circuit is referred to the output, since the OTA is used to generate a signal in a neuron, rather than processing one. 
We use a frequency range from 1~Hz to 1~kHz, corresponding to the typical speed of neuronal processes \cite{rolls10}. In terms of the neuromorphic application, a typical neuron that integrates e.g. a presynaptic waveform generated by the OTA constitutes a low-pass filter with cutoff frequency below 300Hz. The memristors targeted for stimulation by the OTA also exhibit low-pass filter characteristics with cutoff below 1~kHz \cite{cederstroem13a}.

It is expected that flicker noise dominates the other possible noise sources, due to the low frequencies used \cite{allen2000}.
Shot noise and thermal noise are expected to be of lesser significance.
A rough hand calculation (not shown) supports this assumption, resulting in only $17.2$~$\mu V$ total RMS noise voltage for thermal and shot noise in a frequency range from 1~Hz to 1~kHz.

Flicker noise is expected to be dominated by the output transistors  M14-M17 and the diodes M10, M13, since all input and internal signals, including noise, are attenuated by the high current mirror scaling factor.
Different modeling approaches for flicker noise exist \cite{scholten03}, from which a simplified model as used in \cite{rieger10} is most suitable for a noise hand calculation.
This model shows that the noise current amplitude is proportional to the transconductance of the transistor, which in weak inversion is strongly dependent on the transistor's operating point.
In turn, the noise level can be reduced by lowering the gate-source voltage.
The flicker noise coefficient contained in the model in \cite{rieger10} is dependent on the chosen technology and on the transistor dimensions, so that it needs to be determined in simulation.
Thus, a hand calculation of flicker noise can not be done quantitatively without simulation.
We therefore moved directly to noise simulations of the whole circuit to determine overall noise performance.
For room temperature (300~K), nominal simulation yields 391~$\mu V_\mathrm{RMS}$. For a reasonable die operating temperature of 353~K,  simulation yields 925~$\mu V_\mathrm{RMS}$. Comparing this value to the estimated shot and thermal noise amplitudes confirms that those can be neglected compared to the flicker noise contribution.

\begin{figure}
\centering
\includegraphics[width=3.0in]{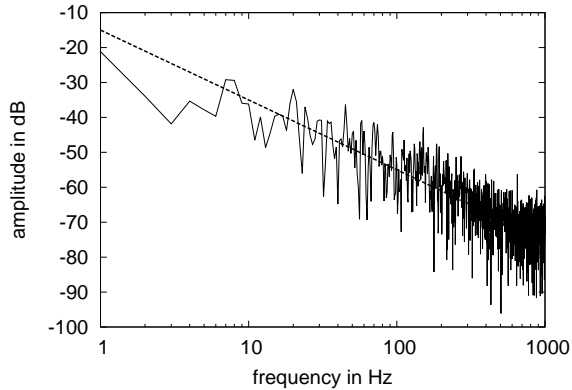}
\caption{Measured spectrum of the OTA noise. The dashed line shows a fitted 1/f noise spectrum.}
\label{fig_noise_spectrum}
\end{figure}

Noise measurements of the manufactured circuit were conducted with an INA103 low noise instrumentation amplifier and an HP54602B oscilloscope.
The RMS noise of the OTA was determined by subtracting the noise of the measurement structure from the result of the complete measurement chain (including the on-chip buffer shown on the right hand side of Fig. \ref{fig_ota_c}).
The resulting spectrum is shown in Fig. \ref{fig_noise_spectrum}, exhibiting a clear 1/f characteristic.
This confirms that flicker noise is indeed the dominant noise contribution.
From the spectrum, an RMS output noise amplitude of 950~$\mu V$ was calculated for the OTA, which agrees well with the noise simulation at 353~K.

\section{Results}
\label{sec_results}

\begin{figure}
\centering
\includegraphics[width=3.0in]{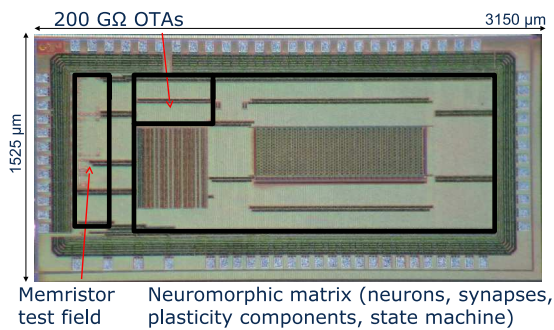}
\caption{Chip photograph of the neuromorphic IC containing the proposed OTA circuit.}
\label{fig_sim}
\end{figure}

The presented OTA circuit was fabricated in a 180~nm CMOS process using 1.8~V supply voltage and 1.8~V core (i.e. digital) transistors. The OTA circuits constitute a separate test block in a neuromorphic chip (see 200G$\Omega$ OTA marker in Fig. \ref{fig_sim}). The chip is otherwise dedicated to a neuron and synapse matrix implementing a novel plasticity rule \cite{mayr10b}. On the left hand side, the chip also contains test structures for CMOS integration of the memristors presented in \cite{mayr12b}. Due to the metal fill, only some of the power rails in the top metal layer are visible.
In order to verify the parameter variations, 13 chips containing 5 identical instances of the OTA have been measured. All measurements in this paper have been carried out at an $I_{bias}$ setting of 1~$\mu$A and $N_{CScale}=\frac{1}{900}$, i.e. for a nominal $g_{m, total}$ of 5~pS.

\begin{figure}
\centering
\includegraphics[width=3.0in]{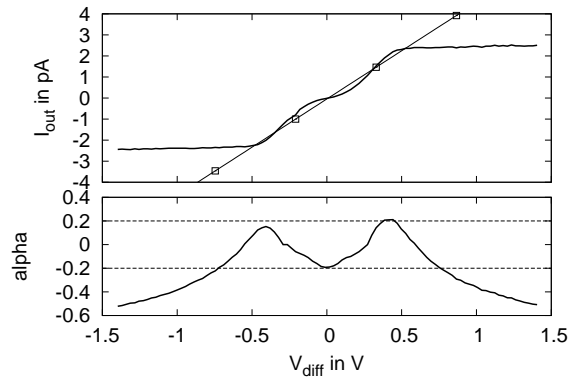}
\caption{a) Measured I-V characteristic of the proposed OTA. b) Measure of linearity, with 20\% range denoted (Measure $\alpha$ as defined in \cite{arnaud03}).}
\label{fig_ui}
\end{figure}

The I-V characteristic of the OTA is depicted in Fig. \ref{fig_ui}. The nonlinearity of the source degeneration near the origin of the diagram prevents the application of relative errors less than 20 percent for the calculation of the linear range. Thus, with a 20 \% error a voltage swing of \unit{1.2}{\volt} can be achieved.

\begin{figure}
\centering
\includegraphics[width=3.0in]{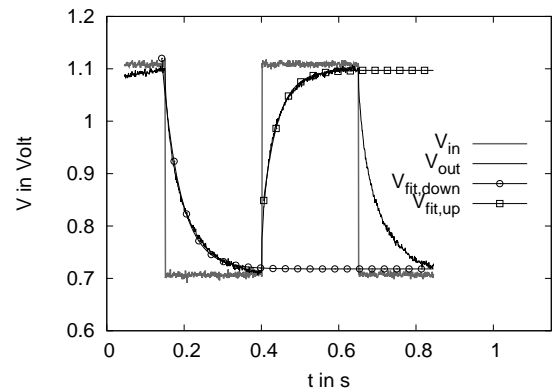}
\caption{Measured exponential decay generated by OTA-C circuit ($\mathrm{V_{out}}$), which was stimulated by a square wave signal at the input ($\mathrm{V_{in}}$). The traces have been fitted with exponential curves $\mathrm{V_{fit,down}}$ and $\mathrm{V_{fit,up}}$ for the falling and rising periods respectively.}
\label{fig_rc}
\end{figure}

As stated the circuit is intended to generate exponentially decaying voltage traces with long time constants. Fig. \ref{fig_rc} shows such a curve stimulated by a square wave at the input of the OTA. The exponential fit indicates that the nonlinearity has no significant influence on the waveform of the decay. In this case the load capacitance is about 250~fF. With a $gm$ of 5~pS as mentioned before time constants on the order of 50~ms can be achieved. 

For both the memristor driver application as well as the waveform generation for the novel learning rule, deviations from the ideal exponential curve shape due to the non-linearity result in corresponding changes of the plasticity curves \cite{mayr10b,mayr12b}. From a computational \cite{pfister06} as well as a biological viewpoint \cite{bi98}, these deviations are not critical. 
When using the OTA as a PSC generator, the PSC is usually integrated on the neuron membrane to generate a charge increase on the membrane capacitance or extend the influence of the presynaptic pulse via the exponentially decaying waveform \cite{rachmuth11,noack10}. Again, some deviation from an ideal exponential waveform are not critical.

The measured output noise of the OTA is 950~$\mu V_\mathrm{RMS}$ in a frequency band of 1~Hz to 1~kHz. In terms of the target neuromorphic application, e.g. employing the OTA for driving a PSC on a neuron membrane with a 1.2~V swing, this noise allows about a 10~bit amplitude resolution level. A similar argument holds with respect to noise when using the OTA as driver for the weight modification of a memristor.

\begin{figure}
\centering
\includegraphics[width=3.0in]{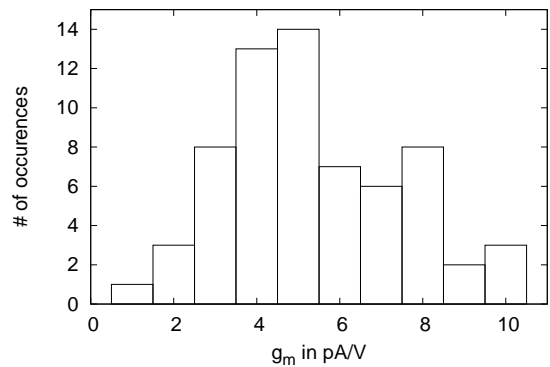}
\caption{Histogram of measured transconductances of 65 instances, mean 5.9 pS, standard deviation 2.1 pS.}
\label{fig_gm}
\end{figure}

\begin{figure}
\centering
\includegraphics[width=3.0in]{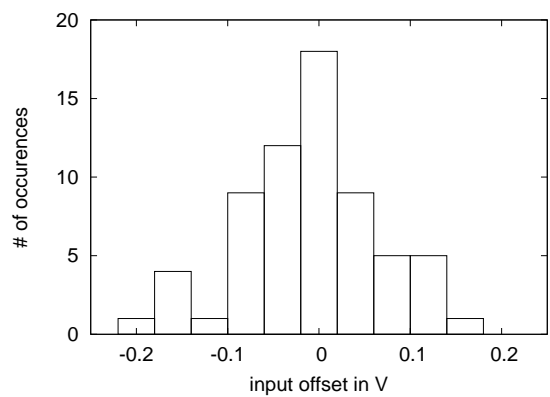}
\caption{Histogram of measured input referred offsets of 65 instances, mean 9.9 mV, standard deviation 75 mV.}
\label{fig_offset}
\end{figure}

Variation of transconductance is depicted in Fig. \ref{fig_gm}, showing a maximum transconductance of about 10~pS. We carried out a monte carlo analysis that shows that the main contributor to the transconductance deviations are the current source transistors (M5--M7) and the output current mirror (M16, M17) followed by the series-parallel current mirrors. The influence of differential pair, current splitting and source degeneration is small in comparison. The same is true for deviations of the offset current shown in Fig. \ref{fig_offset}. Thus, no significant penalty is paid for cascading several gm scaling techniques in series, offset and gm variation is largely determined by circuit parts that are common to many OTAs. The analysis shows the advantage of keeping the current splitter in supra-threshold operation. As the series parallel current mirrors rank only third in contributed error, this also shows the efficacy of the unit transistor approach.

The offset is not critical for the memristor application, since these exhibit a threshold characteristic that is robust to offset \cite{cederstroem13a}. For the novel plasticity rule, the offset has more significance, as it would constitute a continuous weight drift \cite{mayr10a}. In some cases, this drift may be desirable, helping to establish robust synaptic weights \cite{Indiveri2006}. In case where the offset needs to be better controlled, a OTA-wise adjustment may be necessary. The neuromorphic matrix also contains a 2-bit offset correction for OTAs similar to the offset current adjustment of \cite{rieger04}. At the moment, this is not integrated in the low-gm OTA discussed here, adding it would result in additional area of ca. 130~$\mu m^2$. 

In contrast to the offset, the gm variation is significant for both the plasticity rule and the memristors, as it defines the time window of weight modification. We wanted to test the gm mismatch performance in an uncalibrated design and therefore did not include a circuit for OTA-wise adjustment of $I_{bias}$ in the current design. In some cases, the time window variations may not be critical \cite{pfister06}. In applications where it is critical, 4-bit binary weighted current mirrors currently used in the neuromorphic matrix of the IC in Fig. \ref{fig_sim} could be added to the OTA. If applied to the bias current of the OTA, the $gm$ could be individually adjusted for each OTA with area penalty of 190~$\mu m^2$ and gm variation improvement on the order of 4 bit, i.e. to $\pm$ 15\%. Alternatively, too strongly deviating individual circuits could be used only for those parts of the network where they are not critical by applying judicious restructuring of the network \cite{mayr07a}.

Tab. \ref{tab:comparison} gives a comparison of the presented results with those of recently published circuits. Please note: The comparison is somewhat skewed, as in other examples, area and power has been spent on linearity, while the presented OTA achieves an aggressive area and gm scaling at the cost of linearity. Overall, the comparison reflects this design target. As intended for a large scale, high density integration, the OTA discussed here consumes a factor 15 less area than the nearest published comparison and has one of the lowest transconductances, but it needs to assume a very tolerant linearity measure for its linear range entry in table \ref{tab:comparison}. The area comparison is valid even across technologies, as analog circuit area does not scale well with the technology node.

The measured OTA noise level of 950~$\mu V_\mathrm{RMS}$ translates to a noise density of 30~$\mu$V$/\sqrt{\mathrm{Hz}}$ which is comparable with the output noise of \cite{villegas11} with 32~$\mu$V$/\sqrt{\mathrm{Hz}}$, \cite{rieger10} with 72~$\mu$V$/\sqrt{\mathrm{Hz}}$ and \cite{arnaud06} with 56~$\mu$V$/\sqrt{\mathrm{Hz}}$. Due to the above-threshold operation of most of the circuit, the power consumption at 2.7~$\mu$W is somewhat high, but still within the comparison range.

 In relation to the operating voltage, the proposed OTA has the highest voltage swing, which is necessary especially for driving memristors \cite{ohno11,ou13}. In comparison, while the circuit presented in \cite{villegas11} provides an even lower transconductance of 0.5~pS, it suffers from a limited input voltage swing and needs an auxiliary amplifier, which increases the required chip area. In \cite{rieger04}, another method for achieving low gm is presented, using cascaded gm stages and global feedback. However, as all stages operate in weak inversion, the gm variations (not shown in that paper) are probably even more substantial than the ones reported here. Also, stability may be a problem in that design, as there is no local feedback.   

\begin{table*}
\centering
\caption{Comparison with measured performances of low-gm circuits in current literature. If the transconductance is tunable, the maximum range is given. Dependent on the specific reference, the linear range is given for a maximum total harmonic distortion (THD) or with respect to the linearity measure $\alpha$ of \cite{arnaud03}.}
\label{tab:comparison}
\begin{tabular}{p{1cm}p{1.3cm}p{1.3cm}p{1.0cm}p{1.0cm}p{1.2cm}p{1.3cm}p{1.3cm}p{1.2cm}p{2.1cm}}
\toprule
Ref. & Process (\micro\meter) & Area (\milli\square\meter) & Power (\micro\watt) & VDD (\volt) & Linear \linebreak Range (\volt) & Linearity (\% THD or $\alpha$) & Transcon\-ductance range (\pico\siemens) & Noise Density $\left(\frac{\micro\volt}{\surd Hz}\right)$& Application Field\\ 
\midrule

\cite{arnaud06}   & 0.8    & 0.09    & 0.113  & 2     & 0.32   & 5 ($\alpha$)  & 33-35 & 56 & implantable electronics\\ 	  
\cite{chou10}     & 0.18   & $\sim 0.04$      & 2      & 1     & --    & --   & 36000 & 6e-3& biomedical readout\\  
\cite{villegas11} & 0.35   & 0.07    & 0.005  & 1     & 0.14   & 1 (THD)   & 0.5-23e3 & 32& low frequency filter\\    
\cite{rieger10}     & 0.35   & 0.011   & 440     & 5  & 0.5     & 0.2 (THD)    & 75 & 72& nerve signal recording\\     
\cite{huang09}    & 0.35   & 0.046   & 3200   & 5     & 2.6    & 1 (THD)   & 30-25e6 & 0.45 & low-frequency oscillator\\     
This work         & 0.18   & 0.0007  & 2.7 & 1.8   & 1.2 & 20 ($\alpha$) & 5-50 & 30 & neuromorphic signal generation\\ 
\bottomrule
\end{tabular}
\end{table*}

\section{Conclusion}
\label{sec_conc}

This paper presents an OTA with a very low transconductance and small footprint for use in highly-integrated neuromorphic circuits. To achieve this transconductance per area ratio, the circuit employs a combination of previously reported circuit techniques such as current splitting, source degeneration and series-parallel current mirrors. Through the combination of esp. the current scaling techniques, a large part of the circuit operates in strong inversion while the output still achieves the pA output currents in line with a single-digit pS transconductance. The combination of 5~pS transconductance and 700~$\mu m^2$ circuit area is unique in the literature. Especially when regarding the technology node, the reported transconductance is the lowest, i.e. usually such small conductances are only reported for larger nodes. Even in technologies $>$350~nm, only a handful of lower transconductances are reported. The consequent usage of unity transistors in an optimum-matching grid layout keeps circuit variation controlled to some extent despite the agressive overall sizing (see Fig. \ref{fig_gm}). As discussed, this variation can be further downscaled with programmable current mirrors. 

Capacitance per area (e.g. of metal-metal caps) stays virtually constant across different technology nodes. Thus, if one wants to explore even smaller technology nodes with neuromorphic circuits based on the gm-C technique \cite{koickal07}, this kind of agressive OTA shrinkage has to become commonplace since the capacitance area will not shrink, leaving the OTA circuit as the only possibility for area reduction. Ideally, as presented for the circuit here, a combination of downscaling of the transconductance and of the OTA circuit area has to be achieved. Lowered transconductance allows shrinking of the gm-C capacitance for a given time constant, achieving a balanced circuit area reduction of both OTA and attendant capacitance for a target gm-C stage.

While the OTA presented here is targeted at neuromorphic waveform generation, it is by no means relegated to that. The low area requirements and large time constants achievable could be especially interesting for direct integration in dense submicron pixel cells \cite{henker07}. For example, \cite{huang09} presents a voltage-controlled oscillator (VCO) based on one of the OTAs discussed in table \ref{tab:comparison}. In \cite{huang09b}, the same authors employ a VCO with similar characteristics for an oscillating pixel cell. However, presumably for area reasons, in this application the authors choose a VCO based on an inverter chain \cite{scholze11a}, even though an OTA-based VCO would deliver significantly better linearity in terms of voltage-frequency dependence. Thus, a sufficiently small OTA could form a versatile building block in a preprocessing imager matrix, providing pixel-level intensity to frequency conversion \cite{huang09b}, or filtering and feature computation \cite{koenig02}. 

\section*{Acknowledgment}
This research has received funding from the European Union Seventh Framework Programme (FP7/2007- 2013) under grant agreement no. 269459 (CORONET)

\section*{References}
\bibliographystyle{elsarticle-num}
\bibliography{low_gm_ota}
\end{document}